\documentclass[twocolumn,superscriptaddress,nofootinbib,floatfix,aps,prb,citeautoscript,longbibliography,nobalancelastpage,10pt]{revtex4-2}
\usepackage[english]{babel}
\usepackage[utf8]{inputenc}
\usepackage[T1]{fontenc}
\usepackage{lineno}
\usepackage{siunitx}
\usepackage{graphicx}
\usepackage{xcolor}
\usepackage{array}
\usepackage{dcolumn}
\usepackage{bm}
\usepackage{multirow}
\usepackage[version=4]{mhchem}
\usepackage{booktabs}
\usepackage{hyperref}
\usepackage{cleveref}
\usepackage{amsmath,amssymb}  
\usepackage{placeins} 
\usepackage{balance}
\usepackage{float}
\newcommand{\qv}{\bm{q}}
\newcommand{\Rv}{\bm{R}}
\newcommand{\uv}{\bm{u}}

\renewcommand{\Re}{\operatorname{Re}}

\newcommand{\Dif}{\mathrm{D}}
\newcommand{\nequip}{\textsc{NequIP}}
\newcommand{\phonopy}{\textsc{phonopy}}
\newcommand{\angstrom}{\text{\AA}}

\newcommand{\bochum}{Research Center Future Energy Materials and Systems of the University Alliance Ruhr and ICAMS, Ruhr University Bochum, Universit\"atsstra{\ss}e 150, D-44801 Bochum, Germany}
\begin{document}

\title{Linear response for machine-learned interatomic potentials: phonons at arbitrary wave vectors from the primitive cell}

\author{Thalis H. B. da Silva}
\affiliation{\bochum}
\author{Hai-Chen Wang}
\affiliation{\bochum}
\author{Miguel A. L. Marques}
\affiliation{\bochum}

\date{\today}  

\begin{abstract}
Phonon calculations with machine-learned interatomic potentials are almost exclusively performed with finite displacements in supercells, an approach that is exact only at wave vectors commensurate with the supercell and whose error depends on the range of the force constants. In message-passing networks this range grows with the number of layers, and reaches \SI{140}{\angstrom} for the deepest universal models. Here we show that the dynamical matrix of a message-passing potential can be obtained exactly in the primitive cell, at any wave vector, as the linear response of the forces to a displacement modulated by a Bloch phase, in the spirit of density-functional perturbation theory. The phase enters only where the features of neighboring atoms are gathered, and the trained model is used without modification. One further derivative yields group velocities and mode Gr\"uneisen parameters analytically. We apply the method to \nequip{} and Prophet, and reproduce the Hessians of supercells to machine precision. Using these results as a reference, we find that the supercells commonly employed in high-throughput calculations lead to errors of several tenths of a THz in the phonon frequencies, and that these errors decrease slowly with the size of the supercell, in particular for deep models. For a low-symmetry silicate, the supercell predicts a dynamical instability that the model does not have. Quantities that depend on derivatives of the dispersion, such as the group velocities and the Gr\"uneisen parameter at low temperature, are considerably more affected than the frequencies themselves. When both approaches are used to obtain the same force constants, their computational cost is comparable for crystals with symmetry.
\end{abstract}

\maketitle

\section{Introduction}
\label{sec:intro}

Lattice vibrations determine a large part of the thermodynamic and transport properties of solids. Knowing the phonon dispersion therefore gives access to the dynamical stability of a crystal, to its free energy and thermal expansion, and to its lattice thermal conductivity. The same vibrations also couple to the electrons, and in conventional superconductors this coupling mediates the formation of Cooper pairs. All these properties require the dynamical matrix $D(\qv)$ at wave vectors $\qv$ that span the Brillouin zone. In density-functional theory, two routes to $D(\qv)$ are well established. In the finite-displacement, or frozen-phonon, approach~\cite{kresse1995,parlinski1997,phonopy,togo2023}, atoms are displaced in a supercell, the interatomic force constants are obtained from the resulting forces, and $D(\qv)$ follows from a Fourier interpolation. This approach is simple and independent of the underlying code, but it is exact only at wave vectors commensurate with the supercell, and its cost grows with the range of the force constants. Density-functional perturbation theory (DFPT)~\cite{baroni1987,giannozzi1991,gonze1997,dfpt} instead compute the linear response to a perturbation of arbitrary wave vector $\qv$ directly in the primitive cell, at a cost independent of $\qv$. Furthermore, it extends naturally to higher orders, which gives access to anharmonic properties without numerical differentiation~\cite{gonze1989,debernardi1995}.

Machine-learned interatomic potentials (MLIPs)~\cite{behler2007,bartok2010,nequip, batatia2022mace} have changed the scale at which such calculations can be performed. Universal models~\cite{chen2022,deng2023,batatia2023foundation,yang2024mattersim,fu2025esen,wood2025uma} trained on large databases of density-functional calculations now yield phonon frequencies and the derived thermodynamic quantities for thousands of compounds with a good accuracy~\cite{loew2025}, and are being assessed for anharmonic properties such as the thermal conductivity~\cite{pota2026}. Phonon calculations with MLIPs are, however, almost exclusively performed with finite displacements in supercells, a workflow inherited from density-functional theory. This choice has a consequence that is often overlooked. Most current MLIPs are message-passing networks, in which the energy of an atom depends on all atoms within a distance $L r_\text{c}$, where $L$ is the number of layers and $r_\text{c}$ the cutoff radius. The force constants therefore extend up to $2 L r_\text{c}$~\cite{langer2026}, which can range roughly from \SI{20}{\angstrom} to \SI{140}{\angstrom} for common choices of $L$ and $r_\text{c}$. A supercell that contains this range is seldom affordable. In practice, smaller supercells are used, resulting in truncated force constants, but the error is usually not controlled.

A supercell is, however, not required. For classical interatomic potentials, the dynamical matrix has long been evaluated analytically at arbitrary $\qv$ from phased sums of the second derivatives of the energy~\cite{bornhuang,gale1997,gale2003}, and its derivatives with respect to strain were obtained in the same way for the minimization of the free energy~\cite{kantorovich1995,gale1998}. For MLIPs, automatic differentiation gives the exact Hessian of the energy~\cite{gonnheimer2025,koker2026}, and its sparsity can be exploited to reduce the cost~\cite{langer2026}. Fang \textit{et al.}~\cite{fang2024} showed that $D(\qv)$ of a message-passing network can be obtained without a supercell, by differentiating the energy of one cell with respect to the positions of all atoms in its receptive field and summing the result with the appropriate phases. Nondiagonal supercells~\cite{lloydwilliams2015,wang2025ares} reduce the cost of the finite-displacement approach at commensurate wave vectors. All these approaches build the force constants in real space, so that their cost and memory requirements grow with the volume of the receptive field. To the best of our knowledge, an approach that works at fixed $\qv$ in the primitive cell, in the spirit of DFPT, has not been proposed for MLIPs.

In this work, we show that the dynamical matrix of a message-passing MLIP can be obtained exactly from the primitive cell, for any $\qv$. We consider a displacement of the atoms modulated by a Bloch phase and obtain the first-order change of the forces, i.e. the product of the dynamical matrix with this displacement, from a single forward-over-reverse pass of automatic differentiation. The phase enters in one place only, where the features of neighboring atoms are gathered. The trained model is used without retraining, and only small changes to a few of its operations are needed. Since $\qv$ and the strain are ordinary inputs of this calculation, one further derivative yields group velocities and mode Gr\"uneisen parameters analytically.

The remainder of this article is organized as follows. In Sec.~\ref{sec:method} we derive the equations for the dynamical matrix and for its derivatives with respect to $\qv$ and to strain, and discuss their implementation for \nequip~\cite{nequip} and for Prophet~\cite{prophet}. In Sec.~\ref{sec:results} we validate the method against Hessians of supercells, and use it to assess the supercell approach for phonon dispersions, mode Gr\"uneisen parameters, the thermodynamic Gr\"uneisen parameter at low temperature, and group velocities. The advantages and limitations of the method are discussed in Sec.~\ref{sec:discussion}, and our conclusions are presented in Sec.~\ref{sec:conclusions}.

\section{Method}
\label{sec:method}

\subsection{Bloch response of the forces}
\label{sec:formulation}

We consider a crystal with $N$ atoms in the primitive cell, at positions $\bm{\tau}_\kappa + \Rv$, where $\Rv$ is a lattice vector. The interatomic force constants are the second derivatives of the energy $E$ with respect to the atomic displacements $\uv_{\kappa\Rv}$,
\begin{equation}
  \Phi_{\kappa\alpha,\kappa'\beta}(\Rv) = \frac{\partial^2 E}{\partial u_{\kappa\alpha\bm{0}}\,\partial u_{\kappa'\beta\Rv}}\,,
\end{equation}
and the dynamical matrix reads $D_{\kappa\alpha,\kappa'\beta}(\qv) = M_{\kappa\alpha,\kappa'\beta}(\qv)/\sqrt{m_\kappa m_{\kappa'}}$, with $m_\kappa$ the atomic masses and
\begin{equation}
  M_{\kappa\alpha,\kappa'\beta}(\qv) = \sum_{\Rv} \Phi_{\kappa\alpha,\kappa'\beta}(\Rv)\, e^{i\qv\cdot\Rv}\,.
  \label{eq:mq}
\end{equation}
In the supercell approach, the sum in \cref{eq:mq} is restricted to the lattice vectors of the supercell. We proceed differently, and note that the product of $M(\qv)$ with a vector $\uv$ is
\begin{equation}
  \sum_{\kappa'\beta} M_{\kappa\alpha,\kappa'\beta}(\qv)\, u_{\kappa'\beta}
  = - \left.\frac{\Dif F_{\kappa\alpha\bm{0}}}{\Dif\lambda}\right|_{\lambda=0}\,,
  \label{eq:response}
\end{equation}
where $F_{\kappa\alpha\bm{0}}$ is the force on atom $\kappa$ of the cell at the origin, and the derivative is taken along the displacement $\uv_{\kappa\Rv} = \lambda\, \uv_\kappa\, e^{i\qv\cdot\Rv}$ of all atoms of the crystal. The dynamical matrix is therefore the first-order response of the forces to a perturbation modulated by a Bloch phase, exactly as in DFPT~\cite{dfpt}. As the unperturbed crystal is periodic, the first-order change of every quantity attached to an atom, or to a pair of atoms, has the same modulation, i.e., it is multiplied by $e^{i\qv\cdot\Rv}$ under a translation by $\Rv$. It is thus sufficient to know these first-order quantities in the primitive cell.

In a message-passing network, the features $h_i^{(l)}$ of atom $i$ in layer $l$ are updated from the features of its neighbors and from the vectors that connect them. For a crystal this is evaluated on the graph of the primitive cell, where an edge $e$ connects a sender atom $s(e)$ to a receiver atom $t(e)$ and carries the lattice vector $\Rv_e$ of the cell where the sender is located. The only operation of the network that connects different cells is the gather, which copies a quantity $x$ of the sender to the edge,
\begin{equation}
  y_e = x_{s(e)}\,.
\end{equation}
All other operations act on a single atom or on a single edge. The quantity $x$ can be a feature or the position of the atom, as the vector of an edge is $\bm{r}_e = \bm{\tau}_{s(e)} + \Rv_e - \bm{\tau}_{t(e)}$. The first-order change of $x$ in the cell $\Rv_e$ differs from the one in the cell at the origin by the Bloch phase, so that the first-order change $\dot{y}$ of the gathered quantity is
\begin{equation}
  \dot{y}_e = e^{i\phi_e}\, \dot{x}_{s(e)}\,, \qquad \phi_e = \qv\cdot\Rv_e\,.
  \label{eq:gather}
\end{equation}
The forces are obtained by reverse-mode differentiation of the energy, which propagates the adjoints $\bar{x} = \partial E/\partial x$ from the energy back to the positions. The reverse of the gather is a scatter, $\bar{x}_j = \sum_{e:\,s(e)=j} \bar{y}_e$. In the crystal, the atom $j$ of the cell at the origin sends to the receiver of edge $e$ located in the cell $-\Rv_e$, and the first-order change of the adjoint therefore acquires the conjugate phase,
\begin{equation}
  \dot{\bar{x}}_j = \sum_{e:\,s(e)=j} e^{-i\phi_e}\, \dot{\bar{y}}_e\,.
  \label{eq:scatter}
\end{equation}
\Cref{eq:gather,eq:scatter} are the only changes with respect to an ordinary calculation in the primitive cell. The left-hand side of \cref{eq:response} is then obtained by one pass of forward-mode differentiation over the reverse-mode calculation of the forces, with the perturbation $\uv$ at the positions. 
Across the $L$ layers of the forward and the $L$ layers of the reverse pass, a $\uv$ travels along paths of up to $2L$ edges and accumulates their phases. The calculation in the primitive cell thus performs the sum of \cref{eq:mq} implicitly, over the full range $2Lr_\text{c}$ of the force constants, without a supercell. The complete matrix $M(\qv)$ requires $3N$ such passes, each with the cost of a few force evaluations in the primitive cell, for any $\qv$, commensurate or not, and independently of the range of the force constants. No truncation is involved, and the result is the exact dynamical matrix of the model, up to rounding errors.

\subsection{Derivatives with respect to wave vector and strain}
\label{sec:derivatives}

The wave vector enters the calculation only through the phases $\phi_e$, and a homogeneous strain $\eta$ only through the atomic positions and the lattice vectors $\Rv_e$. Both are therefore ordinary differentiable inputs. For two vectors $\uv$ and $\bm{v}$, the scalar $\Re\,[\bm{v}^\dagger M(\qv,\eta)\,\uv]$ is obtained from \cref{eq:response}, and one reverse pass over this calculation yields its gradient with respect to the components of $\qv$ and of $\eta$. We keep the reduced coordinates of the atoms and of $\qv$ fixed under strain, so that a mode is followed as the lattice is deformed.

For a nondegenerate mode of frequency $\omega$ and eigenvector $\bm{e}$ of $D(\qv)$, the Hellmann--Feynman theorem gives $\partial\omega^2/\partial\lambda = \bm{e}^\dagger (\partial D/\partial\lambda)\, \bm{e}$ for any parameter $\lambda$. The group velocity and the mode Gr\"uneisen tensor are then
\begin{align}
  \bm{v}_\text{g} &= \frac{\partial\omega}{\partial\qv} = \frac{1}{2\omega}\, \bm{e}^\dagger \frac{\partial D}{\partial\qv}\, \bm{e}\,, \\
  \gamma_{\alpha\beta} &= -\frac{\partial\ln\omega}{\partial\eta_{\alpha\beta}} = -\frac{1}{2\omega^2}\, \bm{e}^\dagger \frac{\partial D}{\partial\eta_{\alpha\beta}}\, \bm{e}\,,
\end{align}
and the usual mode Gr\"uneisen parameter for an isotropic change of volume is $\gamma = -\partial\ln\omega/\partial\ln V = \frac{1}{3}\sum_\alpha \gamma_{\alpha\alpha}$. These derivatives are analytic, requiring no finite step in $\qv$ or in $\eta$, and hen
ce no matching of modes between the calculations of a finite difference. For a degenerate multiplet, the Hellmann--Feynman expression does not apply directly. As in degenerate perturbation theory, we then evaluate $\partial D/\partial\lambda$ between all eigenvectors of the multiplet and diagonalize the resulting matrix, whose eigenvalues are the derivatives $\partial\omega^2/\partial\lambda$ of the individual modes.

\subsection{Implementation}
\label{sec:impl}

We implemented the method in \textsc{PyTorch}. The gather of \cref{eq:gather} and the scatter of \cref{eq:scatter} are written as two differentiation rules, each being the reverse of the other. The networks that we consider are real, so that we represent a complex first order change by two replicas of the graph of the primitive cell, one for the real and one for the imaginary part, and the multiplication by the phase becomes a rotation between the two. The trained model is called without any change to its code. To this end, the index of the senders is passed to the model as an object that applies \cref{eq:gather} whenever it is used to gather a quantity. In this way we applied the method to \nequip~\cite{nequip}, an equivariant message-passing network built with \textsc{e3nn}~\cite{e3nn}, and to Prophet~\cite{prophet}. The only modifications that were required concern three operations whose library routines do not support, or do not evaluate correctly, the additional reverse pass of \cref{sec:derivatives}. These are the length and the normalization of the edge vectors, and the layer normalization of Prophet, which we wrote explicitly in terms of elementary operations. The calculation of $M(\qv)$ itself does not require them.

Several wave vectors and several columns of $M(\qv)$ can be computed in a single pass, by collecting the corresponding copies of the primitive cell in one disconnected graph. This is relevant in practice, as a graphics processing unit is far from being saturated by a calculation in a primitive cell with a few atoms.

The $3N$ passes at a given wave vector are not all independent. An operation of the space group that maps $\qv$ onto itself, or onto $-\qv$, relates the response to a displacement to the response to its image, so that only the displacements that are inequivalent under these operations have to be calculated. As in DFPT, the method can be used in two ways. The dynamical matrix can be calculated directly at the wave vectors of interest, or it can be calculated at the irreducible wave vectors of a regular grid, rotated to the complete grid, and Fourier interpolated. The latter contains exactly the same information as the none-truncated force constants of the supercell that is commensurate with the grid.

\section{Results}
\label{sec:results}

\subsection{Computational details}
\label{sec:details}

We used three models. The \nequip{} models combine a good accuracy with a very low computational cost, which makes them a natural choice for high-throughput calculations of phonons. Prophet is, at the time of writing, the model with the highest combined performance score in the Matbench Discovery benchmark~\cite{riebesell2025}, which includes the thermal conductivity, and represents the deep networks with a very long formal range of the force constants. NequIP-OAM-S and NequIP-OAM-M~\cite{oam} have two and four layers and cutoff radii of \SI{4.5}{\angstrom} and \SI{6}{\angstrom}, so that their force constants extend to \SI{18}{\angstrom} and \SI{48}{\angstrom}, respectively. 
Prophet-OAME-MBD~\cite{prophet} has ten layers and cutoff radii of \SI{3}{\angstrom} and \SI{7}{\angstrom}, which leads to a formal range of \SI{140}{\angstrom}. Unless stated otherwise, the results were obtained with NequIP-OAM-M.

All calculations were performed in double precision, with the exception of the largest supercells for Prophet, which were calculated in single precision on the central processing unit. We considered silicon, diamond, \ce{MgB2}, bulk and monolayer \ce{MoS2}, and kaolinite, \ce{Al2Si2O5(OH)4}, a layered silicate with 17 atoms in the primitive cell and no symmetry besides the translations (space group $P1$). All structures are at the experimental lattice parameters~\cite{crc97,nagamatsu2001} and with the internal coordinates relaxed with each model.

As a reference we used two supercell calculations. In the first, the force constants were obtained with finite displacements of \SI{0.01}{\angstrom} using \phonopy~\cite{phonopy,togo2023}. In the second, the rows of the Hessian of the supercell were computed exactly by automatic differentiation, which removes the error of the finite displacement and leaves only the truncation of the force constants. In both cases $D(\qv)$ was then obtained by the Fourier interpolation of \phonopy. All timings were measured on a single NVIDIA RTX 4060 Ti graphics card with \SI{16}{\giga\byte} of memory.

\subsection{Validation}
\label{sec:validation}

We first verified that the method reproduces the Hessian of a supercell. For a wave vector commensurate with the supercell, $M(\qv)$ can be assembled without approximation from the rows of the Hessian of the supercell obtained by automatic differentiation. We performed this test for a cell of silicon with displaced atoms and a sheared lattice, so that no element of the matrix vanishes by symmetry, and for wave vectors commensurate with supercells of up to 48 atoms. The largest difference between the two matrices is \SI{2e-13}{\electronvolt\per\angstrom\squared} for NequIP-OAM-S, \SI{2e-14}{\electronvolt\per\angstrom\squared} for NequIP-OAM-M, and \SI{3e-14}{\electronvolt\per\angstrom\squared} for Prophet, while the largest elements of $M(\qv)$ are 10 to \SI{17}{\electronvolt\per\angstrom\squared}. For NequIP-OAM-M and the six materials of this work, the frequencies at the commensurate wave vectors of the paths of \cref{fig:dispersions} agree with those of the supercell Hessian to \SI{e-6}{\tera\hertz}. Note that for Prophet the largest supercell that we could differentiate in double precision had 12 atoms. At an incommensurate wave vector, where no such reference exists, $M(\qv)$ matrices of the \nequip{} models are Hermitian to the same precision, and at $\Gamma$ they fulfill the acoustic sum rule. 

The derivatives of \cref{sec:derivatives} were validated in the same way. The analytic group velocities of silicon agree with finite differences of the frequencies with respect to $\qv$ to \SI{4e-5}{\angstrom\per\pico\second}. Furthermore, in the limit of long wavelengths the velocities of the acoustic modes have to reproduce the solutions of the Christoffel equation with the elastic constants of the model. We obtained the latter from the derivative of the stress with respect to strain, including the relaxation of the internal coordinates, and the two sets of velocities agree to a relative difference of \num{e-7}.

\subsection{Phonon dispersions}
\label{sec:bands}

\begin{figure*}
  \centering
  \includegraphics[width=\textwidth]{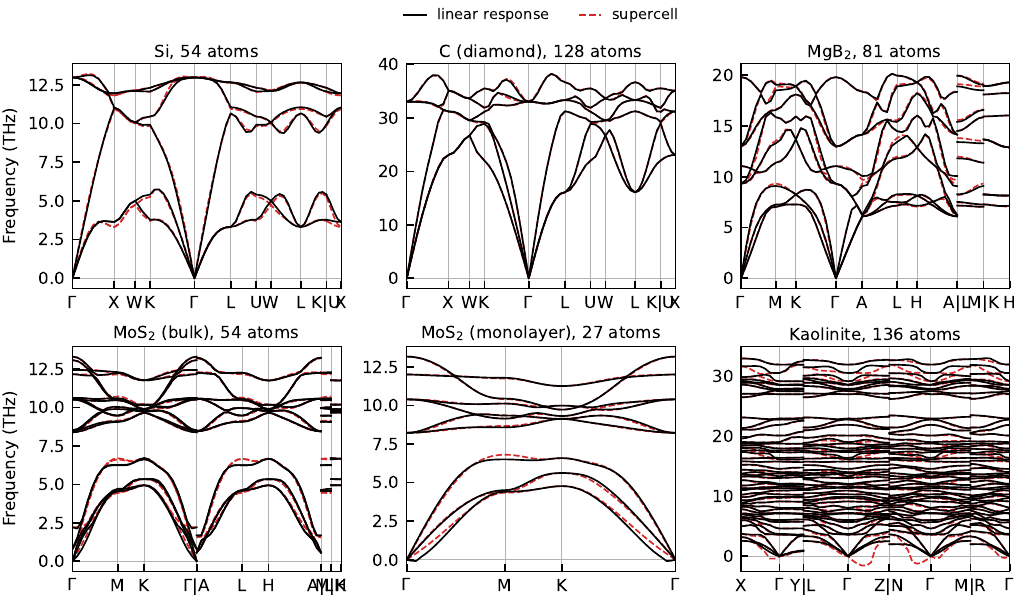}
  \caption{Phonon dispersions calculated with NequIP-OAM-M from the linear response in the primitive cell (black) and from finite displacements in a supercell (red, dashed), with the number of atoms of the supercell indicated in each panel. Imaginary frequencies are shown as negative. For kaolinite we do not show the four O--H stretching modes, which lie around \SI{110}{\tera\hertz}.}
  \label{fig:dispersions}
\end{figure*}

\begin{table}[tbh!]
  \caption{Maximum deviation $\Delta\omega$ between the frequencies of the supercell and of the linear response along the paths of \cref{fig:dispersions}, for NequIP-OAM-M. The last two columns show the time required to obtain the same force constants with the linear response on the grid of wave vectors commensurate with the supercell ($t_\text{LR}$) and with finite displacements in the supercell ($t_\text{FD}$), both using the symmetry of the crystal.}
  \label{tab:dispersions}
  \begin{ruledtabular}
  \begin{tabular}{lcccc}
    & Atoms & $\Delta\omega$ (THz) & $t_\text{LR}$ (s) & $t_\text{FD}$ (s) \\
    \hline
    Si                     & 54  & 0.34 & 0.34 & 0.21 \\
    C (diamond)            & 128 & 0.45 & 0.51 & 0.98 \\
    \ce{MgB2}              & 81  & 0.66 & 0.70 & 0.57 \\
    \ce{MoS2} (bulk)       & 54  & 0.44 & 0.37 & 0.21 \\
    \ce{MoS2} (monolayer)  & 27  & 0.50 & 0.25 & 0.10 \\
    Kaolinite              & 136 & 3.4  & 125  & 22   \\
  \end{tabular}
  \end{ruledtabular}
\end{table}

\begin{figure}
  \centering
  \includegraphics[width=\columnwidth]{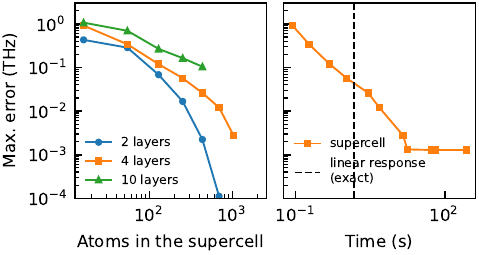}
  \caption{Maximum error of the phonon frequencies of silicon obtained with the supercell approach along the path of \cref{fig:dispersions}. Left: as a function of the number of atoms in the supercell, for models with two (NequIP-OAM-S), four (NequIP-OAM-M), and ten (Prophet) layers. Right: as a function of the time of the calculation with finite displacements, for NequIP-OAM-M and supercells with 16 to 3456 atoms. The dashed line marks the time required by the linear response for the same 60 wave vectors.}
  \label{fig:convergence}
\end{figure}

In \cref{fig:dispersions} we compare the phonon dispersions obtained from the linear response with those of supercells with 27 to 136 atoms, which are typical sizes for high-throughput calculations of phonons. The two calculations coincide at the wave vectors commensurate with the supercell and deviate in between. For the five simpler materials the maximum deviations, listed in \cref{tab:dispersions}, range from \SI{0.34}{\tera\hertz} for silicon to \SI{0.66}{\tera\hertz} for \ce{MgB2}. For kaolinite, with a $2\times2\times2$ supercell, the error reaches \SI{3.4}{\tera\hertz}. More importantly, the supercell calculation yields an acoustic branch with imaginary frequencies, down to $1.5i$~\si{\tera\hertz}, at a quarter of the wave vectors of the path, while the exact dispersion of the model has no imaginary frequency. The truncation of the force constants can therefore lead to the wrong conclusion that a compound is dynamically unstable. The deviations are essentially the same if the finite displacements are replaced by the exact Hessian of the supercell. The error is therefore due to the truncation of the force constants and not to the finite displacement.

The cost of the two approaches can be compared directly when they yield the same information, i.e., when the linear response is calculated on the grid of wave vectors commensurate with the supercell. The results are shown in \cref{tab:dispersions}. For the five materials with high symmetry the two calculations take a comparable time, as only 4 to 14 passes are required for the complete grid. For kaolinite there is no symmetry to exploit, and the linear response is slower by a factor between five and six. The memory required by the linear response is that of the primitive cell multiplied by the number of perturbations that are calculated in one pass, which can be chosen freely, and does not increase with the size of the grid.

The two approaches differ, however, in how the cost increases with the accuracy required, as shown in \cref{fig:convergence} for silicon. For NequIP-OAM-M the error of the supercell decreases from \SI{0.91}{\tera\hertz} for 16 atoms to \SI{0.003}{\tera\hertz} for 1024 atoms, a calculation that required \SI{14}{\second}. For larger supercells the error does not decrease below \SI{0.001}{\tera\hertz}, which is the error of the finite displacement, while the time increases to several minutes. The linear response yields the 60 wave vectors of the path without error in \SI{1.5}{\second}, i.e., in less than \SI{0.03}{\second} per wave vector on this high-symmetry path, and in \SI{0.04}{\second} at a general wave vector. 

For NequIP-OAM-S model with only two layers, a supercell of moderate size (1024 atoms) does contain the complete range of the force constants. The situation is more severe for Prophet, for which the error decreases only slowly with the size of the supercell (\cref{fig:convergence}). It is still \SI{0.27}{\tera\hertz} with 128 atoms, and \SI{0.10}{\tera\hertz} with 432 atoms. The last of these was the largest cell that we could afford, and required \SI{24}{\giga\byte} of memory in single precision. The linear response for the same 60 wave vectors took \SI{68}{\second} and \SI{7.5}{\giga\byte} in double precision. On equal grids the cost of the two approaches is again comparable. For the $4\times4\times4$ grid the linear response required \SI{7.9}{\second} and \SI{7.5}{\giga\byte}, while the finite displacements in the corresponding supercell of 128 atoms \SI{11.1}{\second} and \SI{13.2}{\giga\byte}.

\subsection{Mode Gr\"uneisen parameters}
\label{sec:gruneisen}

\begin{figure}
  \centering
  \includegraphics[width=\columnwidth]{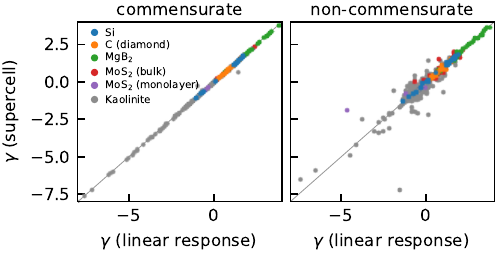}
  \caption{Mode Gr\"uneisen parameters obtained from finite differences of the frequencies of strained supercells against the analytic values of the linear response, for wave vectors commensurate (left) and non-commensurate (right) with the supercell. The supercells are those of \cref{fig:dispersions} and the strain is \SI{0.02}{\percent}. A few modes of kaolinite lie outside the range shown.}
  \label{fig:gruneisen}
\end{figure}

Mode Gr\"uneisen parameters are usually calculated from the frequencies of two or three supercells at different volumes. In \cref{fig:gruneisen} we compare such a calculation, for a strain of \SI{0.02}{\percent} and the supercells of \cref{fig:dispersions}, with the analytic derivatives of \cref{sec:derivatives}. We consider separately the wave vectors that are commensurate with the supercell, where the only error of the supercell calculation is the finite strain, and non-commensurate wave vectors, where the truncation of the force constants also contributes.

At commensurate wave vectors the two calculations coincide. With the exact Hessian of the supercell the largest difference is \num{e-5}, and with finite displacements the mean absolute difference is at most \num{0.0003}, for the five materials with high symmetry. For kaolinite the mean absolute difference is \num{0.006}, as several of its 51 branches are less than \SI{0.02}{\tera\hertz} apart and cross even under this strain. At non-commensurate wave vectors the mean absolute difference amounts to \num{0.02} for diamond, \num{0.03} for silicon, \num{0.04} for \ce{MgB2}, \num{0.09} for bulk \ce{MoS2}, and \num{0.12} for the \ce{MoS2} monolayer, with a maximum of \num{2.8} for the latter. These errors are due to the truncation of the force constants alone. For kaolinite the $2\times2\times2$ supercell fails. Two thirds of the modes deviate by more than \num{0.1}, the mean absolute difference is \num{0.3} for the modes above \SI{5}{\tera\hertz}, and the Gr\"uneisen parameters of the lowest branches are meaningless, as the frequencies of the supercell are close to zero or imaginary.

The strain used above is considerably smaller than what is common in this type of calculation, and this choice turned out to be necessary. With a strain of \SI{0.5}{\percent}, the mean absolute difference at commensurate wave vectors is still below \num{0.02} for silicon, diamond, and \ce{MoS2}, but for \ce{MgB2} it amounts to \num{0.09}, with a maximum of \num{0.89}. These large differences come from two branches that are only \SI{0.03}{\tera\hertz} apart at several wave vectors of the grid, and that have very different Gr\"uneisen parameters. A strain of \SI{0.5}{\percent} shifts one of them, with $\gamma = 3.3$, by \SI{0.8}{\tera\hertz}, so that the two branches cross and the finite difference follows the wrong one. Reducing the strain to \SI{0.1}{\percent} is not sufficient, as the mean absolute difference only decreases to \num{0.06}. The analytic derivative is free of this problem, as it does not require the identification of the modes at different volumes.

The analytic derivative has a different limitation, which we found with NequIP-OAM-S. In silicon at the experimental lattice parameter, the shell of third neighbors lies at \SI{4.503}{\angstrom}, just outside the cutoff radius of \SI{4.5}{\angstrom} of this model. A compression of the lattice brings these atoms inside the cutoff, and the derivatives of the frequencies with respect to strain are not continuous at this point. The analytic derivative and the finite difference then measure different quantities, and the thermodynamic Gr\"uneisen parameter at \SI{10}{\kelvin}, defined below, is \num{-1.13} with the former and \num{-5.85} with the latter. This is a deficiency of the model and not of either method, but it shows that derivatives of this order probe the smoothness of the potential at the cutoff.

\subsection{Thermodynamic Gr\"uneisen parameter at low temperature}
\label{sec:lowT}

\begin{figure}
  \centering
  \includegraphics[width=\columnwidth]{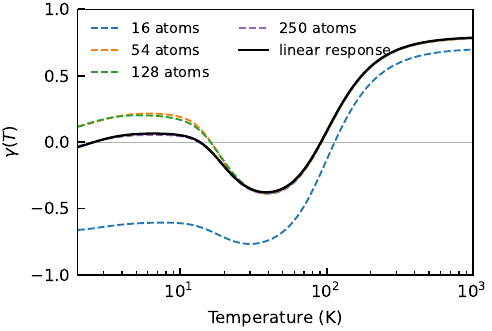}
  \caption{Thermodynamic Gr\"uneisen parameter of silicon as a function of temperature, calculated with NequIP-OAM-M from the linear response (black) and from the exact Hessians of supercells with 16 to 250 atoms (dashed).}
  \label{fig:lowT}
\end{figure}

The thermodynamic Gr\"uneisen parameter is the average of the mode Gr\"uneisen parameters weighted by the heat capacity $c_{\qv\nu}$ of each mode,
\begin{equation}
  \gamma(T) = \frac{\sum_{\qv\nu} c_{\qv\nu}(T)\, \gamma_{\qv\nu}}{\sum_{\qv\nu} c_{\qv\nu}(T)}\,,
\end{equation}
and determines the thermal expansion. At low temperature only the acoustic modes of long wavelength contribute, which in a supercell calculation are obtained entirely by interpolation. We calculated $\gamma(T)$ of silicon on a $60\times60\times60$ mesh of wave vectors, with 5216 points in the irreducible wedge of the Brillouin zone, which took 23 minutes with the linear response. The result changes by \num{0.01} at \SI{10}{\kelvin} with respect to a $30\times30\times30$ mesh, and is not converged with respect to the mesh below this temperature.

The comparison with the supercells is shown in \cref{fig:lowT}. At \SI{300}{\kelvin} the supercells with 54 atoms or more are within \num{0.01} of the linear response, which gives \num{0.693}. At \SI{10}{\kelvin}, however, the linear response yields \num{0.05}, while the supercells with 16, 54, and 128 atoms give \num{-0.61}, \num{0.19}, and \num{0.17}, respectively. Supercells with 250 atoms or more are within \num{0.01} of the linear response. A supercell that is adequate at room temperature can therefore be wrong by a factor of three to four at low temperature. The other two approximations of the supercell approach are much less important here, as the finite strain and the finite displacement change $\gamma(T)$ by less than \num{0.01} and \num{0.001}, respectively.

The sign of $\gamma(T)$ is the sign of the thermal expansion coefficient, which allows for a comparison with experiment. Silicon is known to contract upon heating at low temperature~\cite{lyon1977,middelmann2015}. According to the measurements of \citet{lyon1977}, the linear expansion coefficient is positive and very small below \SI{18}{\kelvin}, where it does not exceed \SI{1.4e-9}{\per\kelvin}, negative from \SI{18}{\kelvin} to \SI{125}{\kelvin}, with a minimum of \SI{-4.7e-7}{\per\kelvin} close to \SI{78}{\kelvin}, and positive above. More recent interferometric measurements~\cite{middelmann2015} place the upper change of sign at \SI{124}{\kelvin}. NequIP-OAM-M reproduces the three regimes, as $\gamma(T)$ calculated with the linear response is positive and small around \SI{10}{\kelvin}, negative between \SI{14}{\kelvin} and \SI{91}{\kelvin}, and positive above. The supercell with 16 atoms misses the first regime entirely, as it yields a negative $\gamma(T)$ down to the lowest temperatures, while those with 54 and 128 atoms overestimate it by a factor of three to four (see \cref{fig:lowT}).

\subsection{Group velocities and the elastic limit}
\label{sec:gvel}

\begin{figure}
  \centering
  \includegraphics[width=\columnwidth]{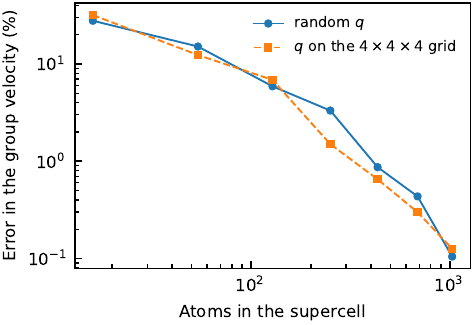}
  \caption{Relative root-mean-square error of the group velocities of silicon obtained from the exact Hessian of a supercell, as a function of the number of atoms in the supercell, for random wave vectors and for the wave vectors of the $4\times4\times4$ grid. The reference is the analytic derivative of the linear response with NequIP-OAM-M.}
  \label{fig:velocities}
\end{figure}

Group velocities enter the lattice thermal conductivity, and are usually obtained from the derivative of the Fourier-interpolated dynamical matrix. In \cref{fig:velocities} we show the error of this procedure for silicon. For random wave vectors the relative error amounts to \SI{28}{\percent}, \SI{15}{\percent}, and \SI{5.9}{\percent} for supercells with 16, 54, and 128 atoms, and falls below \SI{1}{\percent} only for 432 atoms. The velocities converge more slowly than the frequencies. This is clearest at the wave vectors of the $4\times4\times4$ grid, which are commensurate with the supercell of 128 atoms. There the frequencies of this supercell are exact, to \SI{e-14}{\tera\hertz}, but the velocities are in error by \SI{6.9}{\percent}, or \SI{0.96}{\angstrom\per\pico\second}. The Fourier interpolation reproduces the values of the dynamical matrix at the commensurate wave vectors, but not its slope.

The same occurs in the limit of long wavelengths, which gives the elastic constants. A fit of the cubic elastic tensor to the sound velocities of the linear response yields $C_{11} = \SI{115.4}{\giga\pascal}$, $C_{12} = \SI{58.1}{\giga\pascal}$, and $C_{44} = \SI{45.5}{\giga\pascal}$, identical to the values obtained from the derivative of the stress with respect to strain. The same fit to the velocities of the supercell with 54 atoms gives \SI{131.0}{\giga\pascal}, \SI{73.1}{\giga\pascal}, and \SI{54.4}{\giga\pascal}, i.e., errors of \SI{14}{\percent} to \SI{26}{\percent}, and with 128 atoms the error of $C_{44}$ is still \SI{7}{\percent}. We note that these are the elastic constants of the model at the experimental lattice parameter, where it has a residual pressure of \SI{0.8}{\giga\pascal}.

\section{Discussion}
\label{sec:discussion}

The main advantage of the method is that it removes the supercell as a convergence parameter. The dynamical matrix is that of the model, at any wave vector, and its accuracy does not depend on the range of the force constants. This is the argument that favors DFPT over finite displacements in density-functional theory, and it is particularly relevant for MLIPs. In a message-passing network the formal range of the force constants grows with the number of layers, and reaches distances that no supercell can contain for the deepest models. As a consequence, the error of the model and the error of the numerical procedure cannot be separated in a supercell calculation, while they are separated by construction in the linear response. The same holds for incommensurate wave vectors, such as those of charge-density waves or of nesting instabilities, which are accessible directly and not through a sequence of increasingly large supercells.

The second advantage is the access to analytic derivatives. Group velocities and mode Gr\"uneisen parameters are obtained without a finite step and without the identification of modes between calculations. The results of \cref{sec:gruneisen} for \ce{MgB2} show that this is not a marginal issue. A finite strain has to be small with respect to the smallest separation between branches divided by their Gr\"uneisen parameters, a quantity that is not known beforehand, and for \ce{MgB2} a strain of \SI{0.1}{\percent} was still too large. The strain cannot be reduced indefinitely either, as the numerical noise of the frequencies is divided by it. For the group velocities, we found that the Fourier interpolation yields errors of several percent even where the frequencies are exact, and that the elastic constants obtained from the long-wavelength limit inherit these errors. Quantities that depend on the slope of the dispersion, such as the lattice thermal conductivity, are therefore more sensitive to the size of the supercell than the dispersion itself.

The method is also simple to apply. The trained model is used as it is, and the only operation that has to be modified is the gather of the features of the neighbors. We applied it to two different architectures without retraining, and with changes to their code that are limited to the three operations mentioned in \cref{sec:impl}. These operations illustrate, however, that higher-order derivatives are not exercised during training or in standard applications, and that an implementation for a new architecture should be validated against a small supercell, as we did here.

Concerning the computational cost, one has to distinguish two situations. If the linear response is calculated on a regular grid and interpolated, it yields the same force constants as the corresponding supercell, at a similar cost when the crystal has symmetry and at a higher cost, by a factor between five and six in our example, when it has none. In our tests a pass of the linear response is in fact 6 to 11 times more expensive than a force evaluation for the same number of atoms, which is compensated by the smaller number of calculations that are required. Any quantity integrated over the Brillouin zone has to be converged with respect to this grid, exactly as it has to be converged with respect to the supercell. The difference is that the linear response can also be evaluated at any other wave vector, which provides a direct measure of the error of the interpolation, and that the grid can be refined without increasing the memory. If, on the other hand, the dynamical matrix is calculated directly at many wave vectors, the cost is proportional to their number, and a dense mesh is expensive. We believe that there is room for improvement, as the forward-over-reverse pass is not compiled and the batching of wave vectors and perturbations was not optimized. We recall that the efficiency of DFPT is also the result of many years of development.

Regarding previous work, the approach of \citet{fang2024} also yields the exact dynamical matrix without a supercell, from the second derivatives of the energy with respect to the positions of all atoms within the receptive field of the primitive cell. The two approaches differ in the same way as the real-space and the reciprocal-space formulations of lattice dynamics. In the former, the cost and the memory are determined by the size of the receptive field, and all wave vectors are then obtained at once. In the latter, they are those of the primitive cell for each wave vector, independently of the receptive field.

\section{Conclusions}
\label{sec:conclusions}

We presented a method to calculate the dynamical matrix of a message-passing interatomic potential in the primitive cell, at an arbitrary wave vector and without approximations. The method follows the logic of density-functional perturbation theory. The first-order response of the forces to a displacement modulated by a Bloch phase is propagated through the network, and the phase enters only in the operation that gathers the features of the neighboring atoms. The trained model is not modified, and we applied the method to two different architectures. As the wave vector and the strain are differentiable inputs, group velocities and mode Gr\"uneisen parameters are obtained analytically, and the symmetry of the crystal is used to reduce the number of perturbations at each wave vector.

Having an exact reference, we could quantify the error of the supercell approach. For a model with four layers and supercells with 27 to 128 atoms, the phonon frequencies of the materials that we considered are in error by \SIrange{0.3}{0.7}{\tera\hertz}, and for a model with ten layers the largest supercell that we could afford still deviates by \SI{0.10}{\tera\hertz}. For kaolinite, a low-symmetry silicate, the supercell yields imaginary frequencies that are absent from the exact dispersion of the model, i.e., a spurious dynamical instability. Quantities that depend on derivatives of the dispersion are more sensitive. The group velocities of silicon obtained with a supercell of 128 atoms are in error by \SI{6}{\percent}, also at the wave vectors where its frequencies are exact, and the thermodynamic Gr\"uneisen parameter at low temperature is overestimated by a factor of three. Furthermore, mode Gr\"uneisen parameters obtained from finite strains can follow the wrong branch when two modes are close in frequency, a problem that does not exist for the analytic derivative. When both approaches are used to obtain the same force constants, their cost is similar for crystals with high symmetry.

The natural extension of this work is the third order, which gives access to the phonon lifetimes and to the lattice thermal conductivity. The three-phonon matrix elements are the second-order response of the forces to two displacements modulated by the wave vectors $\qv$ and $\qv'$. In a message-passing network this requires that the gather multiplies each of the two first-order tangents by its own phase, and the mixed second-order change by the phase of $\qv + \qv'$. The rest of the calculation is again the unmodified model, and we verified for a model potential that this construction reproduces the third derivatives obtained from an explicit supercell. As in density-functional perturbation theory~\cite{gonze1989,debernardi1995,paulatto2013}, the third-order matrices can then be calculated on a coarse grid of pairs of wave vectors and Fourier interpolated, and we expect that the range of the third-order force constants will be as relevant for the convergence of the thermal conductivity as the range of the second-order ones is for the quantities studied here. We leave this for future work.

\section{Data availability}
\label{sec:data_ava}
Not applicable.

\section{Code availability}
\label{sec:code_ava}
he code will be made available at \url{https://github.com/hyllios/utils/tree/main/mlips-analytic-derivatives}.

\section{Acknowledgements}
T.H.B.S. and M.A.L.M. were supported by a collaboration between the Kavli Foundation, Klaus Tschira Stiftung, and Kevin Wells, as part of the SuperC collaboration. M.A.L.M. was supported by the Simons Foundation through the Collaboration on New Frontiers in Superconductivity (Grant No. SFI-MPS-NFS-00006741-10). Part of the calculations were performed on the HPC cluster Elysium of the Ruhr University Bochum, subsidised by the DFG (INST 213/1055-1).

\section{Author Contributions}
T.H.B.S. and H.-C.W. developed the method, wrote the code, and performed the tests. M.A.L.M. conceived and directed the research. All authors contributed to the interpretation of the results and to the writing of the manuscript.

\section{Competing Interests}
The authors declare that they have no competing interests.

\bibliography{bib.bib}

@article{nequip,
  author  = {Batzner, Simon and Musaelian, Albert and Sun, Lixin and Geiger, Mario and Mailoa, Jonathan P. and Kornbluth, Mordechai and Molinari, Nicola and Smidt, Tess E. and Kozinsky, Boris},
  title   = {E(3)-equivariant graph neural networks for data-efficient and accurate interatomic potentials},
  journal = {Nat. Commun.},
  volume  = {13},
  pages   = {2453},
  year    = {2022},
  doi     = {10.1038/s41467-022-29939-5}
}

@article{dfpt,
  author  = {Baroni, Stefano and de Gironcoli, Stefano and Dal Corso, Andrea and Giannozzi, Paolo},
  title   = {Phonons and related crystal properties from density-functional perturbation theory},
  journal = {Rev. Mod. Phys.},
  volume  = {73},
  pages   = {515--562},
  year    = {2001},
  doi     = {10.1103/RevModPhys.73.515}
}

@article{phonopy,
  author  = {Togo, Atsushi and Tanaka, Isao},
  title   = {First principles phonon calculations in materials science},
  journal = {Scr. Mater.},
  volume  = {108},
  pages   = {1--5},
  year    = {2015},
  doi     = {10.1016/j.scriptamat.2015.07.021}
}

@misc{e3nn,
  author  = {Geiger, Mario and Smidt, Tess},
  title   = {e3nn: {E}uclidean neural networks},
  year    = {2022},
  eprint  = {2207.09453},
  archivePrefix = {arXiv}
}

@misc{oam,
  author  = {Kavanagh, Se{\'a}n R. and {MIR Group}},
  title   = {{NequIP} \& {Allegro} foundation potentials},
  howpublished = {Zenodo},
  year    = {2026},
  doi     = {10.5281/zenodo.18775904},
  note    = {doi:10.5281/zenodo.18775904}
}

@article{loew2025,
  author  = {Loew, Antoine and Sun, Dewen and Wang, Hai-Chen and Botti, Silvana and Marques, Miguel A. L.},
  title   = {Universal machine learning interatomic potentials are ready for phonons},
  journal = {npj Comput. Mater.},
  volume  = {11},
  pages   = {178},
  year    = {2025},
  doi     = {10.1038/s41524-025-01650-1}
}

@book{bornhuang,
  author    = {Born, Max and Huang, Kun},
  title     = {Dynamical Theory of Crystal Lattices},
  publisher = {Oxford University Press},
  address   = {Oxford},
  year      = {1954}
}

@misc{fang2024,
  author  = {Fang, Shiang and Geiger, Mario and Checkelsky, Joseph G. and Smidt, Tess},
  title   = {Phonon predictions with {E}(3)-equivariant graph neural networks},
  year    = {2024},
  eprint  = {2403.11347},
  archivePrefix = {arXiv}
}

@article{gale1997,
  author  = {Gale, Julian D.},
  title   = {{GULP}: A computer program for the symmetry-adapted simulation of solids},
  journal = {J. Chem. Soc., Faraday Trans.},
  volume  = {93},
  pages   = {629--637},
  year    = {1997},
  doi     = {10.1039/a606455h}
}

@article{gale2003,
  author  = {Gale, Julian D. and Rohl, Andrew L.},
  title   = {The General Utility Lattice Program ({GULP})},
  journal = {Mol. Simul.},
  volume  = {29},
  pages   = {291--341},
  year    = {2003},
  doi     = {10.1080/0892702031000104887}
}

@article{gonnheimer2025,
  author  = {G{\"o}nnheimer, Nils and Reuter, Karsten and Margraf, Johannes T.},
  title   = {Beyond numerical {H}essians: Higher-order derivatives for machine learning interatomic potentials via automatic differentiation},
  journal = {J. Chem. Theory Comput.},
  volume  = {21},
  pages   = {4742--4752},
  year    = {2025},
  doi     = {10.1021/acs.jctc.4c01790}
}

@misc{koker2026,
  author  = {Koker, Teddy and Gangan, Abhijeet and Kotak, Mit and Marian, Jaime and Smidt, Tess},
  title   = {{PFT}: Phonon fine-tuning for machine learned interatomic potentials},
  year    = {2026},
  eprint  = {2601.07742},
  archivePrefix = {arXiv}
}

@misc{langer2026,
  author  = {Langer, Marcel F. and Hill, Adrian and Ceriotti, Michele},
  title   = {Truncated automatic sparse differentiation for machine learning interatomic potentials},
  year    = {2026},
  eprint  = {2609.20510},
  archivePrefix = {arXiv}
}

@article{lloydwilliams2015,
  author  = {Lloyd-Williams, Jonathan H. and Monserrat, Bartomeu},
  title   = {Lattice dynamics and electron-phonon coupling calculations using nondiagonal supercells},
  journal = {Phys. Rev. B},
  volume  = {92},
  pages   = {184301},
  year    = {2015},
  doi     = {10.1103/PhysRevB.92.184301}
}

@misc{wang2025ares,
  author  = {Wang, Qian and Li, Jiaxiang and Xie, Yu},
  title   = {{ARES-Phonon}: Phonon calculation package using nondiagonal supercell finite displacement method with machine learning},
  year    = {2025},
  eprint  = {2503.11013},
  archivePrefix = {arXiv}
}

@article{behler2007,
  author  = {Behler, J{\"o}rg and Parrinello, Michele},
  title   = {Generalized neural-network representation of high-dimensional potential-energy surfaces},
  journal = {Phys. Rev. Lett.},
  volume  = {98},
  pages   = {146401},
  year    = {2007},
  doi     = {10.1103/PhysRevLett.98.146401}
}

@article{bartok2010,
  author  = {Bart{\'o}k, Albert P. and Payne, Mike C. and Kondor, Risi and Cs{\'a}nyi, G{\'a}bor},
  title   = {Gaussian approximation potentials: The accuracy of quantum mechanics, without the electrons},
  journal = {Phys. Rev. Lett.},
  volume  = {104},
  pages   = {136403},
  year    = {2010},
  doi     = {10.1103/PhysRevLett.104.136403}
}

@inproceedings{batatia2022mace,
  author    = {Batatia, Ilyes and Kov{\'a}cs, D{\'a}vid P{\'e}ter and Simm, Gregor N. C. and Ortner, Christoph and Cs{\'a}nyi, G{\'a}bor},
  title     = {{MACE}: Higher order equivariant message passing neural networks for fast and accurate force fields},
  booktitle = {Advances in Neural Information Processing Systems},
  volume    = {35},
  pages     = {11423--11436},
  year      = {2022},
  doi       = {10.52202/068431-0830}
}

@article{chen2022,
  author  = {Chen, Chi and Ong, Shyue Ping},
  title   = {A universal graph deep learning interatomic potential for the periodic table},
  journal = {Nat. Comput. Sci.},
  volume  = {2},
  pages   = {718--728},
  year    = {2022},
  doi     = {10.1038/s43588-022-00349-3}
}

@article{deng2023,
  author  = {Deng, Bowen and Zhong, Peichen and Jun, KyuJung and Riebesell, Janosh and Han, Kevin and Bartel, Christopher J. and Ceder, Gerbrand},
  title   = {{CHGNet} as a pretrained universal neural network potential for charge-informed atomistic modelling},
  journal = {Nat. Mach. Intell.},
  volume  = {5},
  pages   = {1031--1041},
  year    = {2023},
  doi     = {10.1038/s42256-023-00716-3}
}

@article{batatia2023foundation,
  author  = {Batatia, Ilyes and Benner, Philipp and Chiang, Yuan and Elena, Alin M. and Kov{\'a}cs, D{\'a}vid P. and Riebesell, Janosh and others},
  title   = {A foundation model for atomistic materials chemistry},
  journal = {J. Chem. Phys.},
  volume  = {163},
  pages   = {184110},
  year    = {2025},
  doi     = {10.1063/5.0297006}
}

@misc{yang2024mattersim,
  author  = {Yang, Han and Hu, Chenxi and Zhou, Yichi and Liu, Xixian and Shi, Yu and Li, Jielan and others},
  title   = {{MatterSim}: A deep learning atomistic model across elements, temperatures and pressures},
  year    = {2024},
  eprint  = {2405.04967},
  archivePrefix = {arXiv}
}

@misc{fu2025esen,
  author  = {Fu, Xiang and Wood, Brandon M. and Barroso-Luque, Luis and Levine, Daniel S. and Gao, Meng and Dzamba, Misko and Zitnick, C. Lawrence},
  title   = {Learning smooth and expressive interatomic potentials for physical property prediction},
  year    = {2025},
  eprint  = {2502.12147},
  archivePrefix = {arXiv}
}

@inproceedings{wood2025uma,
  author    = {Wood, Brandon M. and Dzamba, Misko and Fu, Xiang and Gao, Meng and Shuaibi, Muhammed and Barroso-Luque, Luis and others},
  title     = {{UMA}: A family of universal models for atoms},
  booktitle = {Advances in Neural Information Processing Systems},
  volume    = {38},
  pages     = {143528--143564},
  year      = {2025},
  doi       = {10.52202/085713-4310}
}

@article{baroni1987,
  author  = {Baroni, Stefano and Giannozzi, Paolo and Testa, Andrea},
  title   = {Green's-function approach to linear response in solids},
  journal = {Phys. Rev. Lett.},
  volume  = {58},
  pages   = {1861--1864},
  year    = {1987},
  doi     = {10.1103/PhysRevLett.58.1861}
}

@article{giannozzi1991,
  author  = {Giannozzi, Paolo and de Gironcoli, Stefano and Pavone, Pasquale and Baroni, Stefano},
  title   = {Ab initio calculation of phonon dispersions in semiconductors},
  journal = {Phys. Rev. B},
  volume  = {43},
  pages   = {7231--7242},
  year    = {1991},
  doi     = {10.1103/PhysRevB.43.7231}
}

@article{gonze1997,
  author  = {Gonze, Xavier and Lee, Changyol},
  title   = {Dynamical matrices, {B}orn effective charges, dielectric permittivity tensors, and interatomic force constants from density-functional perturbation theory},
  journal = {Phys. Rev. B},
  volume  = {55},
  pages   = {10355--10368},
  year    = {1997},
  doi     = {10.1103/PhysRevB.55.10355}
}

@article{gonze1989,
  author  = {Gonze, X. and Vigneron, J.-P.},
  title   = {Density-functional approach to nonlinear-response coefficients of solids},
  journal = {Phys. Rev. B},
  volume  = {39},
  pages   = {13120--13128},
  year    = {1989},
  doi     = {10.1103/PhysRevB.39.13120}
}

@article{debernardi1995,
  author  = {Debernardi, Alberto and Baroni, Stefano and Molinari, Elisa},
  title   = {Anharmonic phonon lifetimes in semiconductors from density-functional perturbation theory},
  journal = {Phys. Rev. Lett.},
  volume  = {75},
  pages   = {1819--1822},
  year    = {1995},
  doi     = {10.1103/PhysRevLett.75.1819}
}

@article{kresse1995,
  author  = {Kresse, G. and Furthm{\"u}ller, J. and Hafner, J.},
  title   = {Ab initio force constant approach to phonon dispersion relations of diamond and graphite},
  journal = {Europhys. Lett.},
  volume  = {32},
  pages   = {729--734},
  year    = {1995},
  doi     = {10.1209/0295-5075/32/9/005}
}

@article{parlinski1997,
  author  = {Parlinski, K. and Li, Z. Q. and Kawazoe, Y.},
  title   = {First-principles determination of the soft mode in cubic {ZrO$_2$}},
  journal = {Phys. Rev. Lett.},
  volume  = {78},
  pages   = {4063--4066},
  year    = {1997},
  doi     = {10.1103/PhysRevLett.78.4063}
}

@article{togo2023,
  author  = {Togo, Atsushi},
  title   = {First-principles phonon calculations with {P}honopy and {P}hono3py},
  journal = {J. Phys. Soc. Jpn.},
  volume  = {92},
  pages   = {012001},
  year    = {2023},
  doi     = {10.7566/JPSJ.92.012001}
}

@article{kantorovich1995,
  author  = {Kantorovich, L. N.},
  title   = {Thermoelastic properties of perfect crystals with nonprimitive lattices. {I}. {G}eneral theory},
  journal = {Phys. Rev. B},
  volume  = {51},
  pages   = {3520--3534},
  year    = {1995},
  doi     = {10.1103/PhysRevB.51.3520}
}

@article{gale1998,
  author  = {Gale, Julian D.},
  title   = {Analytical free energy minimization of silica polymorphs},
  journal = {J. Phys. Chem. B},
  volume  = {102},
  number = {28},
  pages   = {5423--5431},
  year    = {1998},
  doi     = {10.1021/jp980396p}
}

@article{pota2026,
  author  = {P{\'o}ta, Bal{\'a}zs and Ahlawat, Paramvir and Cs{\'a}nyi, G{\'a}bor and Simoncelli, Michele},
  title   = {Thermal conductivity predictions with foundation atomistic models},
  journal = {Nat. Commun.},
  volume  = {17},
  pages   = {10566},
  year    = {2026},
  doi     = {10.1038/s41467-026-76391-w}
}

@misc{prophet,
  author  = {{Kairos Materials}},
  title   = {Prophet: Scaling atomistic foundation models across composition, configuration, and spin},
  howpublished = {\url{https://www.kairosmaterials.com/papers/Prophet.pdf}},
  year    = {2026}
}

@article{riebesell2025,
  author  = {Riebesell, Janosh and Goodall, Rhys E. A. and Benner, Philipp and Chiang, Yuan and Deng, Bowen and Ceder, Gerbrand and Asta, Mark and Lee, Alpha A. and Jain, Anubhav and Persson, Kristin A.},
  title   = {A framework to evaluate machine learning crystal stability predictions},
  journal = {Nat. Mach. Intell.},
  volume  = {7},
  pages   = {836--847},
  year    = {2025},
  doi     = {10.1038/s42256-025-01055-1}
}

@article{nagamatsu2001,
  author  = {Nagamatsu, Jun and Nakagawa, Norimasa and Muranaka, Takahiro and Zenitani, Yuji and Akimitsu, Jun},
  title   = {Superconductivity at 39~{K} in magnesium diboride},
  journal = {Nature},
  volume  = {410},
  pages   = {63--64},
  year    = {2001},
  doi     = {10.1038/35065039}
}

@article{middelmann2015,
  author  = {Middelmann, Thomas and Walkov, Alexander and Bartl, Guido and Sch{\"o}del, Ren{\'e}},
  title   = {Thermal expansion coefficient of single-crystal silicon from 7~{K} to 293~{K}},
  journal = {Phys. Rev. B},
  volume  = {92},
  pages   = {174113},
  year    = {2015},
  doi     = {10.1103/PhysRevB.92.174113}
}

@article{lyon1977,
  author  = {Lyon, K. G. and Salinger, G. L. and Swenson, C. A. and White, G. K.},
  title   = {Linear thermal expansion measurements on silicon from 6 to 340~{K}},
  journal = {J. Appl. Phys.},
  volume  = {48},
  pages   = {865--868},
  year    = {1977},
  doi     = {10.1063/1.323747}
}

@book{crc97,
  editor    = {Haynes, William M.},
  title     = {{CRC} Handbook of Chemistry and Physics},
  edition   = {97},
  publisher = {CRC Press},
  address   = {Boca Raton},
  year      = {2016}
}

@article{paulatto2013,
  author  = {Paulatto, Lorenzo and Mauri, Francesco and Lazzeri, Michele},
  title   = {Anharmonic properties from a generalized third-order \textit{ab initio} approach: Theory and applications to graphite and graphene},
  journal = {Phys. Rev. B},
  volume  = {87},
  pages   = {214303},
  year    = {2013},
  doi     = {10.1103/PhysRevB.87.214303}
}
\end{document}